\documentclass[twocolumn,aps,prb,amsmath,floatfix]{revtex4}
\usepackage{graphicx}
\begin{document}
\title{Fermi-liquid, non-Fermi-liquid, and Mott phases
       in iron pnictides and cuprates} 
%      Fermi-liquid to non-Fermi-liquid transition in iron pnictide 
%      LaFeAsO \\ studied within exact diagonalization dynamical mean field theor}
\author{Hiroshi Ishida$^1$ and Ansgar Liebsch$^2$} 
\affiliation{$^1$College of Humanities and Sciences, Nihon University,~Tokyo 156,
             Japan\\  
             $^2$Institut f\"ur Festk\"orperforschung, 
             Forschungszentrum J\"ulich, 
             52425 J\"ulich, Germany} 
 \begin{abstract}
The role of Coulomb correlations in the iron pnictide LaFeAsO is studied 
by generalizing exact diagonalization dynamical mean field theory to five 
orbitals. For rotationally invariant Hund's rule coupling a transition from a  
paramagnetic Fermi-liquid phase to a non-Fermi-liquid metallic phase exhibiting 
frozen moments is found at moderate Coulomb energies. For Ising-like exchange, 
this transition occurs at a considerably lower critical Coulomb energy. 
The correlation-induced scattering rate as a function of doping relative to 
half-filling, i.e., $\delta=n/5-1$, where $n=6$ for the undoped material, is 
shown to be qualitatively similar to the one in the two-dimensional single-band 
Hubbard model which is commonly used to study strong correlations in high-$T_c$
cuprates. In this scenario, the parent Mott insulator of LaFeAsO is the  
half-filled $n=5$ limit, while the undoped $n=6$ material corresponds to the 
critical doping region $\delta_c\approx0.2$ in the cuprates, on the verge 
between the Fermi-liquid phase of the overdoped region and the non-Fermi-liquid 
pseudogap phase in the underdoped region. 
\\
\mbox{\hskip1cm}  
electron systems 
\end{abstract}
\maketitle

\section{Introduction}

The discovery of superconductivity in the iron-based pnictides 
\cite{kamihira,ren,chen,cruz} has recently led to an intense debate on the 
question whether Coulomb correlations in these materials play a role of similar 
importance as in the high-$T_c$ cuprates.
%---------------------------------------------------------------------
\cite{haule1,si1,kuroki,dago,baskaran,raghu,maier,yildirim,mazin1,wu,mazin2,craco,%
korschunov,miyake,nakamura,shorikov,graser,haule2,hackl,luca,kou,anisimov,veronica,%
laad,yu,kuroki2,yang,aichhorn,skornyakov,qazil,hozoi,si2,arita,daghofer,zhou,mazin3}
%---------------------------------------------------------------------
Moreover, in contrast to the effective single-band character of the
$3d_{x^2-y^2}$ states in the cuprates, the number of relevant $d$ orbitals
in the pnictides has also been a subject of discussion. In addition,   
because of the multi-band character of the Fe $3d$ bands, the interplay 
of Coulomb repulsion and Hund's rule coupling should be of crucial 
importance for the pnictides. Finally, while it is generally accepted 
that the non-Fermi-liquid pseudogap properties in the underdoped regime 
of the cuprates are associated with the vicinity to the Mott insulating 
phase, in the pnictides it is not yet clear whether there exists a nearby
Mott phase in the range of realistic Coulomb and exchange energies.         

To study the effect of Coulomb correlations in iron pnictides, various 
groups\cite{haule1,craco,korschunov,miyake,%
shorikov,haule2,anisimov,aichhorn,skornyakov,arita}
have applied single-site dynamical mean field theory\cite{dmft} (DMFT).
Depending on the details of the single-particle Hamiltonian and the 
magnitude of Coulomb and exchange energies, weakly or strongly  
correlated solutions were found. 
In the present work we extend exact diagonalization\cite{ed,dagotto} (ED) 
DMFT to five orbitals in order to investigate correlations effects in 
LaFeAsO as a function of Coulomb energy. The single-particle properties 
are described in terms of a five-band tight-binding Hamiltonian.\cite{graser}  
Each $d$ orbital hybridizes with two bath levels, giving $15$ levels in total. 
The five baths are coupled indirectly via the interorbital $3d$ Coulomb and 
exchange interactions. Thus, the Hilbert space is extremely large and 
finite-size effects are effectively reduced. The advantage of this multi-orbital 
ED/DMFT approach is that it is particularly useful at low temperatures and 
that it can handle large Coulomb energies and full Hund exchange. 
As will be discussed below, the latter feature is of special importance 
in the pnictides since Hund's rule coupling leads to electronic 
properties that differ qualitatively from those obtained for the more 
approximate Ising-like exchange treatment. The sensitive role of exchange
interactions in the pnictides was also noted in several previous papers. 
\cite{haule1,craco,haule2,aichhorn}  

The main result of this work is the identification of a  
paramagnetic Fermi-liquid to non-Fermi-liquid transition at moderate 
Coulomb energies, $U_c\approx3$~eV ($J=0.75$~eV), i.e., well below the 
overall width of the Fe $3d$ bands, $W\approx 4.5$~eV. This incoherent 
metallic phase extends up to rather large values of $U$ ($>6$~eV if $J$ 
is kept fixed at $0.75$~eV). It is associated with the formation of local 
moments and with substantial low-frequency scattering rates in all $3d$ 
bands. Below this transition, all bands exhibit strong correlation-induced 
effective mass enhancement. These properties are intimately related to 
the multi-band nature of LaFeAsO and the Hund's rule coupling among the 
Fe $3d$ subbands. A similar spin freezing transition had been found 
recenly by Werner {\it et al.}\cite{werner} in a fully degenerate 
three-band model. Within the present five-band system, a Mott phase 
is not obtained below $U=6$~eV. 
Thus, correlation effects in LaFeAsO appear to be related to the formation 
of local moments within the non-Fermi-liquid phase, and not to the vicinity 
of a Mott insulating phase.     

On the other hand, LaFeAsO readily turns into a Mott insulator at realistic 
Coulomb energies in the hypothetical limit of one-hole doping, i.e., $n=5$. 
The $3d$ bands then become half-filled and split into lower and upper Hubbard 
bands. At intermediate hole doping, non-Fermi-liquid behavior dominates, 
while for electron doping ($n>6$) the system becomes a normal Fermi liquid.
These results suggest a remarkable correspondence between the multi-band 
compound LaFeAsO and the two-dimensional single-band Hubbard model. Indeed,
if the correlation-induced scattering rate is plotted as a function of doping 
relative to half-filling, i.e., $\delta=n/5-1$ ($n=6$ for pure LaFeAsO), 
both systems exhibit the same sequence of phases for increasing $\delta$:  
a Mott insulator at half-filling, a non-Fermi-liquid phase up to a critical 
doping of the order of $\delta_c\approx 0.15 \ldots 0.20$, and a weakly 
correlated Fermi liquid for $\delta>\delta_c$. In this scenario, the 
paramagnetic phase diagrams of iron pnictides and cuprates are strikingly 
similar, with LaFeAsO ($\delta=0.2$) located slightly on the overdoped 
side above critical doping. As a result, the system exhibits a clear 
asymmetry with respect to doping. Whereas electron doping beyond $n=6$ 
($\delta>0.2$) reinforces Fermi-liquid properties, hole doping $n<6$ 
($\delta<0.2$) enhances bad-metallic behavior. Although the real material 
is undoubtedly more complex because of doping-dependent single-particle 
properties and the presence of antiferromagnetism\cite{cruz} at $n=6$, 
we believe that the above picture nevertheless provides a useful new 
perspective for the role of correlation effects in iron pnictides in 
comparison with analogous physics in the cuprates. 

If exchange interactions among Fe $3d$ electrons are approximated in 
terms of Ising-like exchange, i.e., by  neglecting spin-flip and 
pair-exchange processes, the Fermi-liquid to non-Fermi-liquid transition 
still exists, but the critical value of $U$ is shifted down to about $2$~eV 
(assuming $J=U/4$) and the transition is more abrupt. Similar qualitative 
changes from Hund to Ising exchange were found previously also for the 
Mott transition in two-band Hubbard models.\cite{pruschke,prl05} 

Because of the strong hybridization between Fe $3d$ and As $4p$ and O $2p$ 
states, there are indications that an accurate Wannier representation should 
encompass not only $d$ but also $p$ type basis functions, even if Coulomb 
correlations are explicitly only included among the $3d$ orbitals.
\cite{anisimov,kuroki2,haule2,aichhorn,miyake,nakamura} 
A particular consequence of $dp$ hybridization is that the effective $3d$ 
Coulomb interaction is considerably reduced. Moreover, Coulomb interactions 
among different $d$ states are differently screened, giving rise to 
nonisotropic intraorbital and interorbital matrix elements.
\cite{nakamura,aichhorn} 
The present approach is general in the sense that these choices only affect 
the single-particle Hamiltonian and not the evaluation of the $d$ electron 
self-energy matrix. Nevertheless, in this initial five-orbital ED/DMFT
study we use, for simplicity, a purely $d$ electron tight-binding picture
\cite{graser} in order to elucidate the nature of the transition from 
Fermi-liquid to non-Fermi-liquid behavior. A more detailed investigation 
within a $dp$ formulation is planned for future work. As in previous papers,
\cite{haule1,craco,korschunov,miyake,shorikov,haule2,anisimov,aichhorn,%
skornyakov,arita} we focus here on the paramagnetic phase.

The outline of this paper is as follows. In Section II we discuss several
theoretical details concerning the single-particle properties of LaFeAsO
and the ED/DMFT procedure that is used to evaluate the Fe $3d$ self-energy
components. In Section III we analyze the results, with particular focus 
on the differences obtained for Hund and Ising exchange.
Subsection A discusses the neutral system, whereas the effect of doping
and the analogy between pnictides and cuprates are the subject of Subsection B.     
Section IV contains the summary.     
     
\section{Multi-band ED/DMFT} 

In this Section we briefly outline the theoretical details of the
multi-band ED/DMFT approach used in this work. The focus is on the 
role of Coulomb correlations within the Fe $3d$ subbands of LaFeAsO. 
The single-particle properties are described in terms of the five-band 
tight-binding Hamiltonian $H({\bf k})$ which was recently derived by 
Graser {\it et al.}\cite{graser} for a single plane of Fe atoms from 
an accurate fit to the density functional results by Cao {\it et al.}\cite{cao} 
The low-energy part of these bands are in excellent agreement
with analogous calculations by Singh {\it et al.}\cite{singh}  
The basis functions are 
$d_{xz}$, $d_{yz}$, $d_{x^2-y^2}$, $d_{xy}$, and $d_{3z^2-r^2}$,
where the $x,\,y$ axes point along Fe nearest neighbor directions.
The first three of these orbitals comprise the $t_{2g}$ subset,
the remaining two orbitals represent the $e_g$ subset. Hopping up 
to fifth nearest neighbors was included in the tight-binding fit.
The onsite energies are 
$E_{xz,yz}=0.13$~eV, $E_{x^2-y^2}=-0.22$~eV,
$E_{xy}=0.3$~eV, and $E_{z^2}=-0.211$~eV. 
Thus the $d_{xz,yz,xy}$ levels lie $0.2\ldots0.4$~eV above the 
$d_{x^2-y^2,z^2}$ levels. The hopping 
parameters are given in the Appendix of Ref.~\cite{graser} and the     
one-electron band structure corresponding to this Hamiltonian
is shown in Fig.~5 of this Reference. 

The Fe $3d$ density of states 
components are shown in Fig.~1. For symmetry reasons, the $d_{xz}$ 
and $d_{yz}$ components are degenerate. The widths of the $t_{2g}$
and $e_g$ bands are approximately $3.0$~eV and $4.0$~eV, respectively.
The total band width is about $4.5$~eV. All Fe $3d$ bands exhibit a 
pronounced bonding-antibonding splitting, with a deep pseudogap at 
small positive energies, due to the hybridization with
the neighboring LaAsO layers. In the absence of correlations, the 
occupancies of these bands are: $n_{xz,yz}=0.58$, $n_{x^2-y^2}=0.53$,
$n_{xy}=0.52$, and $n_{z^2}=0.78$. Note that these occupancies do not
reflect the crystal field splitting among the onsite energies because 
of the complex shape of the density of states components. In particular,
the bands of $d_{z^2}$ character are considerably more occupied than 
those of $d_{x^2-y^2}$ character, despite the fact that 
$E_{z^2}>E_{x^2-y^2}$.  
A similar situation exists in the layer compound Na$_x$CoO$_2$, 
where $n_{e_g'}>n_{a_g}$ although $E_{e_g'}>E_{a_g}$.\cite{NaCo}

% xz           .58178
% yz           .58178
% xx-yy        .52564
% xy           .52127
% zz           .78248

Previously, we have used finite-temperature ED/DMFT to investigate 
correlation effects in $t_{2g}$ three-band transition metal oxides, 
such as Ca$_2$RuO$_4$,\cite{CRO} Na$_x$CoO$_2$,\cite{NaCo,perroni} 
LaTiO$_3$,\cite{LTO} and V$_2$O$_3$.\cite{prb08} It was shown that, in 
these systems, accurate projections of the lattice Green's function 
onto a finite cluster consisting of three impurity levels and six 
bath levels ($2$ bath levels per $t_{2g}$ orbital) can be achieved, 
yielding an overall cluster size $n_s=9$. Since the different baths 
are indirectly coupled via the $3d$ interorbital Coulomb and exchange
interactions, the spacing between excitation energies is rather small,
so that finite-size effects are greatly diminished.    
Here, we generalize this approach to five impurity orbitals, each 
coupled to two bath levels, i.e., $n_s=15$. 

\begin{figure} [t!] %1
\begin{center}
\includegraphics[width=4.5cm,height=6.5cm,angle=-90]{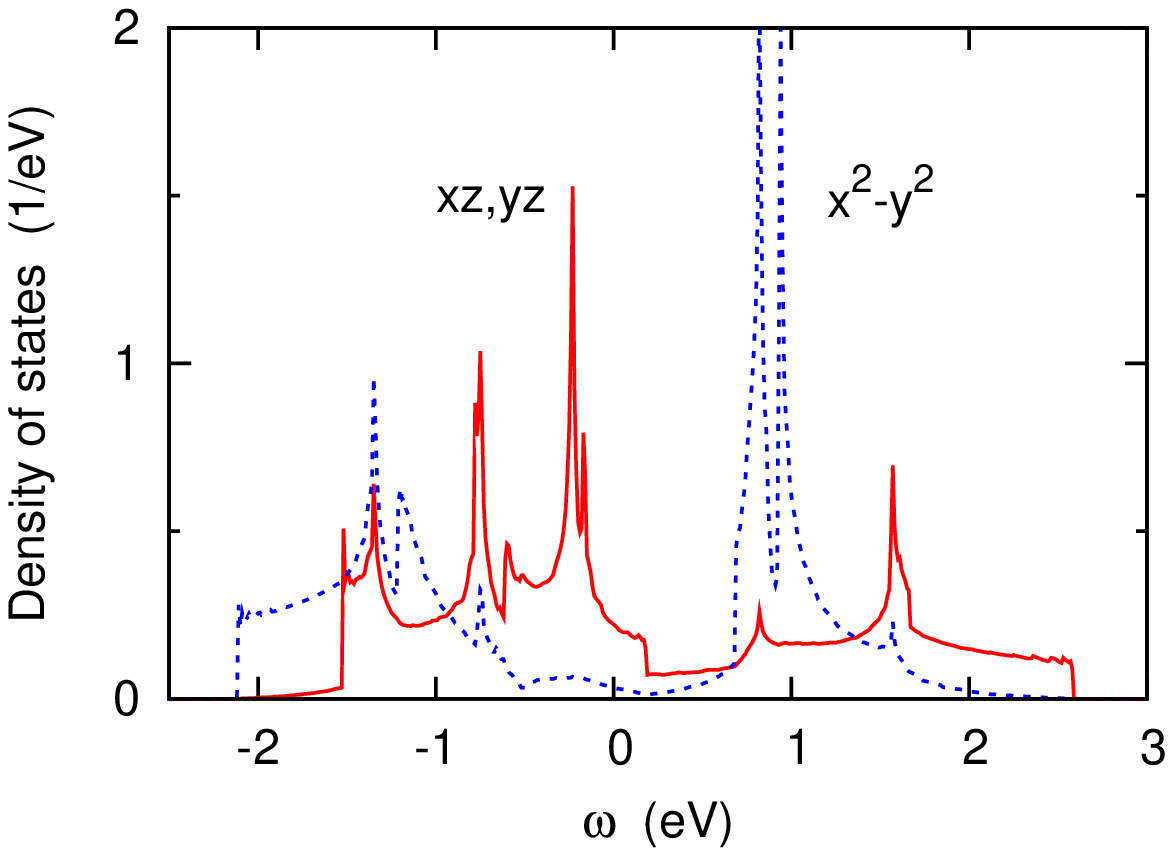}
\end{center}
\vskip-7mm \ \ \ (a)\hfill \mbox{\hskip5mm}
\begin{center}
\includegraphics[width=4.5cm,height=6.5cm,angle=-90]{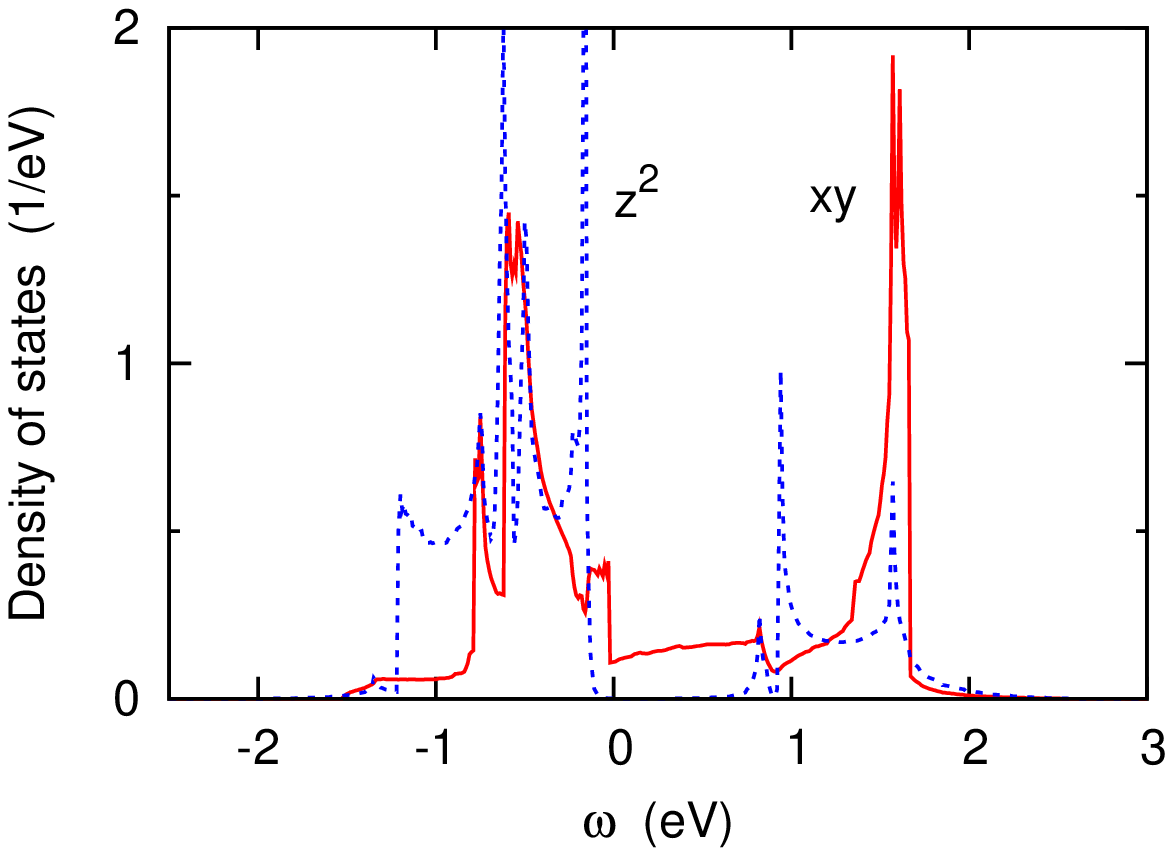}
\end{center}
\vskip-7mm \ \ \ (b)\hfill \mbox{\hskip5mm}
\caption{(Color online)
Density of states components of LaFeAsO for tight-binding
band structure derived in Ref.~\cite{graser}.
(a) $t_{2g}$ components $\rho_{xz,yz}$ and $\rho_{x^2-y^2}$,
(b) $e_{g}$  components $\rho_{xy}$ and $\rho_{z^2}$.  
}\label{dos}\end{figure}

Denoting the tight-binding matrix elements by $H_{mn}({\bf k})$   
the interacting Hamiltonian is given by:
\begin{eqnarray}
   H &=& \sum_{mn{\bf k}\sigma} H_{mn}({\bf k}) c^+_{m{\bf k}\sigma}
                  c_{n{\bf k}\sigma}  \nonumber\\
    && + \sum_{im} U n_{im\uparrow} n_{im\downarrow}  + \!\!\!\!
  \sum_{i m< m'\sigma\sigma'}\!\!\!\! (U'-J\delta_{\sigma\sigma'}) 
                 n_{im\sigma} n_{im'\sigma'}                \nonumber\\
  && -\!\!\!\sum_{im\ne m'}\!\!\! J'[c_{im\uparrow}^+ c_{im\downarrow}
            c_{im'\downarrow}^+ c_{im'\uparrow}   
           + c_{im\uparrow}^+ c_{im\downarrow}^+    
            c_{im'\uparrow} c_{im'\downarrow}]  \nonumber\\
                                               \label{H}
\end{eqnarray}
where $c_{im\sigma}^{(+)}$ are annihilation (creation) operators 
for electrons on site $i$ in orbital $m$ with spin $\sigma$  and
$n_{im\sigma}=c_{im\sigma}^+ c_{im\sigma}$. $c_{{\bf k}m\sigma}^{(+)}$ 
are the corresponding Fourier components. The intra-orbital and 
inter-orbital Coulomb energies are denoted by $U$ and $U'$. The exchange 
integral is $J$, where $U'=U-2J$ because of rotational invariance.
Spin-flip and pair-exchange terms are denoted explicitly by $J'$. 
In the case of isotropic Hund exchange, one has $J'=J$. In the case 
of Ising-like exchange these terms are neglected, i.e., $J'=0$.

The aim of the five-orbital single-site DMFT calculation is to derive 
the local self-energy matrix $\Sigma_{mn}(\omega)$ which describes the 
modification of the single-particle bands due to Coulomb interactions.   
The local lattice Green's function is given by
\begin{equation}
   G_{mn} (i\omega_n)  = \sum_{\bf k} \Big(
   i\omega_n + \mu - H({\bf k})  - \Sigma(i\omega_n)
          \Big)^{-1}_{mn}
                                                   \label{G}
\end{equation}
where $\omega_n=(2n+1)\pi T$ are Matsubara frequencies and $\mu$ 
is the chemical potential. Since we consider paramagnetic systems, 
the spin index of $G$ and $\Sigma$ is omitted. As a result of the 
symmetry properties of $H_{mn}({\bf k})$, the density of states 
matrix is diagonal: $\rho_{mn}(\omega)=\delta_{mn}\rho_{m}(\omega)$. 
Local Coulomb interactions preserve this symmetry, so that $G_{mn}$ 
and $\Sigma_{mn}$ are also diagonal. We denote these 
components by $G_{m}(i\omega_n)$ and $\Sigma_{m}(i\omega_n)$.
We point out that, because of the non-diagonal nature of $H({\bf k})$,
each $G_{m}$ component is influenced by all components $\Sigma_{m}$.

For the purpose of the quantum impurity calculation within DMFT it 
is necessary to first remove the self-energy from the central site. 
This step yields the impurity Green's function     
\begin{equation}
G_{0,m}(i\omega_n)=[G_m(i\omega_n)^{-1}+\Sigma_m(i\omega_n)]^{-1}.
                                                      \label{G0}
\end{equation} 
Within ED/DMFT the lattice impurity Green's function 
$G_{0}$ is approximated in terms of an Anderson impurity model for 
a cluster consisting of impurity levels $\varepsilon_{m=1\ldots5}$ 
and bath levels $\varepsilon_{k=6\ldots 15}$, which are coupled via 
hopping matrix elements $V_{mk}$. Thus, 
$G_{0,m}(i\omega_n) \approx  G_{0,m}^{cl}(i\omega_n)$, where
\begin{equation}
 G_{0,m}^{cl}(i\omega_n) = \Big(i\omega_n + \mu - 
         \varepsilon_m - \sum_{k=6}^{15} 
         \frac{ \vert V_{mk} \vert^2 }
         {i\omega_n - \varepsilon_{k}}\Big)^{-1}  .
                                                       \label{G0cl}
\end{equation} 
Since $G_{0,m}^{cl}$ is diagonal in orbital indices, each impurity 
level couples to its own bath containing two levels: Orbital $1$ couples 
to bath levels $6,7$, orbital $2$ to bath levels {8,9}, etc. Each of the 
four independent functions $G_{0,m}^{cl}(i\omega_n)$ therefore involves 
5 adjustable parameters: one impurity level, two bath levels, and two 
hopping elements. These parameters are found by using a standard 
minimization procedure. The quality of these fits using 5 parameters
is very good, as shown in several previous works.\cite{prb08,prb09} 

\begin{figure}  [t!] %2
\begin{center}
\includegraphics[width=4.5cm,height=6.5cm,angle=-90]{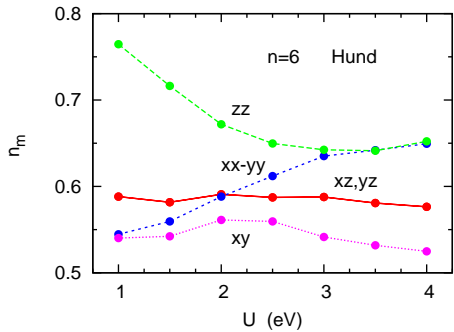}
\end{center}\vskip-3mm
\caption{(Color online)
Fe $3d$ orbital occupancies (per spin) as functions of Coulomb 
interaction for $n=6$, with $J=U/4$ for $U\le3$~eV and $J=0.75$~eV 
for $U>3$~eV; Hund exchange. 
}\label{nvsU}\end{figure}

As a result of the ED quantum impurity calculation one obtains the 
finite temperature cluster Green's function which is also diagonal: 
$G^{cl}_{m}(i\omega_n)$.  In analogy to Eq.~(\ref{G0}) the cluster 
self-energy is given by
\begin{equation} 
\Sigma^{cl}_m(i\omega_n) = G^{cl}_{0,m}(i\omega_n)^{-1} - 
                           G^{cl}_m(i\omega_n)^{-1}. 
                                                \label{sigcl}
\end{equation}
The important physical assumption within DMFT is now that this cluster 
self-energy provides an adequate representation of the self-energy 
of the extended solid, i.e., 
$\Sigma^{cl}_m(i\omega_n) \approx \Sigma_m(i\omega_n)$, which is then 
used in Eq.~(\ref{G}) to derive the lattice Green's function in the next 
iteration step. Further details can be found in Ref.\cite{perroni}. 

Since the cluster Hamiltonian is extremely sparse (typically only 20 to
30 off-diagonal elements per row), the quantum impurity calculation
is conveniently carried out by using the Arnoldi algorithm.\cite{arnoldi} 
The largest spin sector corresponds to $(n_\uparrow,n_\downarrow)=(7,8)$,
giving matrix dimension $(15!/(7!8!))^2= 6435^2$. To reduce storage
requirements, we have rewritten our multi-orbital ED/DMFT code so
that large basis vectors of size $2^{2 n_s}$ are avoided by keeping 
only vectors of size $2^{n_s}$. Moreover, the 
Arnoldi scheme is readily parallelized. Thus, using 32 processors 
the largest Hamiltonian subblock requires less than 1 GB storage. 
Since the spacing between excited states is very small, at finite 
temperatures a large number of states may contribute to the cluster 
Green's function. To reduce computational time in this first five-band 
ED study, we perform the DMFT calculation at $T=0.01$~eV, but retain only 
the lowest few states, making sure that ground state degeneracies are 
properly treated. Using 32 processors, one iteration then takes of the 
order of one to four hours.

\section{Results and discussion}
\subsection{Undoped LaFeAsO}

An important consequence of local Coulomb interactions is the 
rearrangement of electrons among subbands. Fig.~\ref{nvsU} 
shows the variation of the Fe $3d$ orbital occupancies with $U$ at 
$n=6$ total occupancy. The precise values of $U$ and $J$ for LaFeAsO 
depend sensitively on the basis functions used for the $3d$ Hamiltonian.
\cite{anisimov,aichhorn,shorikov,miyake}
In the range $U\le3$~eV we chose $J=U/4$. For illustrative 
purposes we also show results for larger $U$. In this range $J$ is kept 
constant at $0.75$~eV, in order to avoid unrealistically large Hund 
parameters. A charge flow from $d_{z^2}$ to $d_{x^2-y^2}$ 
is seen to take place, thereby reducing the orbital polarization of 
the uncorrelated bands. The occupancies of the $d_{xz,yz,xy}$ orbitals 
are less strongly affected by correlations. The results shown are for 
full Hund coupling. Ising exchange yields a similar charge rearrangement 
predominantly between $d_{z^2}$ and $d_{x^2-y^2}$ orbitals, with only 
slightly larger modifications of the $d_{xz,yz,xy}$ occupancies than
seen in Fig.~\ref{nvsU}. Near $U=2.5$~eV, all $3d$ occupancies are within
about 10~\% of the average occupancy $0.6$. 
The origin of the unusual reduced $d_{z^2,x^2-y^2}$
orbital polarization is the complex bonding-antibonding shape of the 
density of states which yields $n_{z^2}>n_{x^2-y^2}$ although 
$E_{z^2}>E_{x^2-y^2}$. A correlation induced reduction of orbital 
polarization is also found in Na$_x$CoO$_2$ which exhibits a similar
pseudogap in the density of states as a result of the strong $3d-2p$
hybridization in the planar geometry.\cite{NaCo}  
  
\begin{figure}  [t!] %3
\begin{center}
 \includegraphics[width=4.5cm,height=6.5cm,angle=-90]{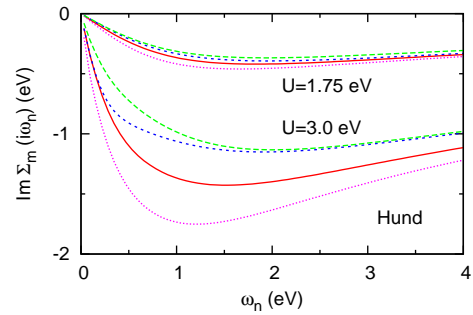}
\end{center}
\vskip-7mm \ \ \ (a)\hfill \mbox{\hskip5mm}
\begin{center}
 \includegraphics[width=4.5cm,height=6.5cm,angle=-90]{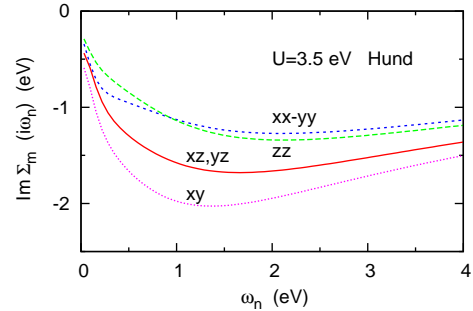}
\end{center}
\vskip-7mm \ \ \ (b)\hfill \mbox{\hskip5mm}
\begin{center}
 \includegraphics[width=4.5cm,height=6.5cm,angle=-90]{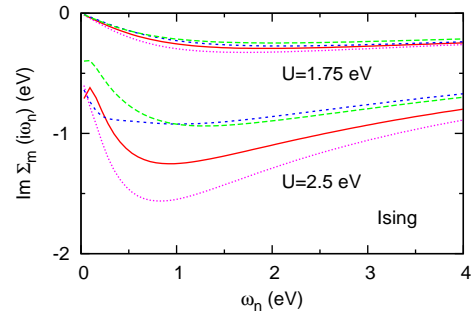}
\end{center}
\vskip-7mm \ \ \ (c)\hfill \mbox{\hskip5mm}
\vskip-1mm
\caption{(Color online)
Imagimary part of self-energy components $\Sigma_m(i\omega_n)$  
as function of Matsubara frequency for several Coulomb energies.
(a) and (b): Hund coupling; (c) Ising exchange. 
The color coding is defined in panel (b).
}\label{sigma}\end{figure}
 
To illustrate the effect of Coulomb correlations on the Fe $3d$ bands in more 
detail, we show in Fig.~\ref{sigma} the self-energy components as functions
of Matsubara frequency. In the case of Hund coupling, the initial slope 
of Im\,$\Sigma_m(i\omega_n)\sim\omega_n$ increases up to about $U=3$~eV, 
until the quasi-particle weights are reduced to about $0.15\ldots 0.3$. 
Beyond this Coulomb energy, the self-energy components exhibit a finite   
onset of $\mbox{-0.3\ldots -0.6}$~eV, indicating that states at the Fermi 
level acquire a finite lifetime. Since this onset is much larger than what is 
expected due to finite temperature, it implies a breakdown of Fermi-liquid
behavior.    
The loss of coherence is strongest for $d_{xy}$ and weakest for $d_{z^2}$.
Ising exchange also gives rise non-Fermi-liquid behavior, except that the 
onset occurs at about $U=2$~eV, i.e., at considerably lower Coulomb energy
than for Hund exchange. 

\begin{figure}  [t!] %4
\begin{center}
\includegraphics[width=4.5cm,height=6.5cm,angle=-90]{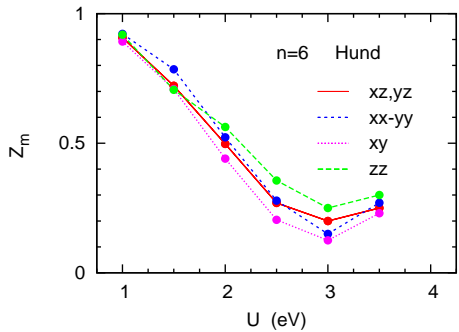}
\end{center}\vskip-3mm
\caption{(Color online)
Quasiparticle weights $Z_m$ of Fe $3d$ orbitals as functions of
Coulomb interaction for $n=6$; Hund exchange. 
}\label{ZvsU}\end{figure}

Fig.~\ref{ZvsU} shows the orbital dependent quasi-particle weights 
$Z_m =1/ [1- {\rm Im} d\Sigma_m(i\omega)/d\omega\vert_{\omega\rightarrow0}]$ 
as functions of Coulomb energy for Hund exchange. Despite the different $3d$ 
orbital occupancies and different $t_{2g}$ and $e_g$ band widths, all five 
bands are seen to exhibit a similar reduction of $Z_m$ with increasing $U$.   
Ising exchange yields a slightly steeper decrease of $Z_m$ up to about 
$U=2$~eV, beyond which all self-energy components show a finite onset.

There exists strong experimental evidence that Fe pnictides exhibit 
an orbital dependent effective mass enhancement, and an concomitant 
narrowing of quasiparticle bands, of about a factor of 2 to 3.
\cite{yang,qazil,lu,ding} According to the results shown in Fig.~\ref{ZvsU}, 
these experimental findings are compatible with the present five-orbital
picture if $U\approx 2.0\ldots2.5$~eV, with $J\approx U/4$ and full Hund's 
rule coupling. The system would then be just below the Fermi-liquid 
to non-Fermi-liquid phase boundary. For Ising exchange, on the other hand,  
these Coulomb and exchange parameters would imply non-Fermi-liquid behavior. 
The above small values of $U$ evidently reflect the fact that, in a purely 
$3d$ electron description, screening via As and O $p$ states greatly reduces 
the Fe $3d$ Coulomb interaction. Correspondingly larger $U$ are appropriate
within a more refined $dp$ description.
\cite{miyake,anisimov,kuroki2,aichhorn,skornyakov}           
  
\begin{figure}  [t] %5
\begin{center}
\includegraphics[width=4.5cm,height=6.5cm,angle=-90]{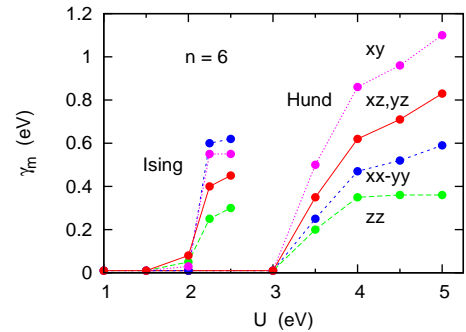}
\end{center}\vskip-3mm
\caption{(Color online)
Orbital dependent low-frequency scattering rates $\gamma_m$ as functions 
of Coulomb interaction for Hund and Ising exchange.
}\label{SigvsU}\end{figure}

Figure~\ref{SigvsU} shows the variation of the low-frequency scattering 
rates $\gamma_m = -{\rm Im}\Sigma_m(i\omega\rightarrow 0)$ with Coulomb
energy. For Hund as well as Ising exchange, the onset of non-Fermi-liquid
occurs for all bands at the same critical $U$.  
Whereas Ising coupling leads to a sudden rise near $U\approx2$~eV, 
Hund exchange shows a smoother increase near $U\approx3$~eV. These 
results are consistent with previous ones for a degenerate two-band model 
\cite{pruschke} which showed that Ising coupling yields a lower 
critical Coulomb energy and a change from a continuous to a first-order  
transition. The damping rates for Ising coupling near $U=2.5$~eV are about 
$0.4\ldots0.7$~eV, in qualitative agreement with quantum Monte Carlo (QMC) 
results in Refs.~\cite{haule1,aichhorn}. For Hund coupling near $U=3.5$~eV 
they are of similar magnitude. They continue to rise at larger $U$ and 
indicate increasing orbital differentiation, with larger damping for 
$d_{xy,xz,yz}$ than for $d_{x^2-y^2,z^2}$.

The transition from coherent to incoherent metallic behavior should also
manifest itself in the temperature variation of the Fe $3d$ self-energy.
This has been studied recently for LaFeAsO$_{1-x}$F$_x$ at $x=0.1$ within  
continuous-time QMC DMFT for full Hund exchange.\cite{haule2} The coherence 
temperature was shown to be a highly sensitive function of $J$, becoming 
extremely small for $J\approx0.7$~eV, which is close to the value assumed 
here ($J=0.75$~eV for $U>3$~eV). 

%The accessibility of the low-temperature coherence region is  limited 
%for computational reasons. Although the present ED DMFT calculations
%are carried out at a rather low temperature ($T=0.01$~eV), the small 
%number of bath levels does not permit to analyze further the coherence 
%region below the first Matsubara frequency, $\omega_0 \approx 0.03$~eV. 

\begin{figure}  [t!] %6
\begin{center}
\includegraphics[width=4.5cm,height=6.5cm,angle=-90]{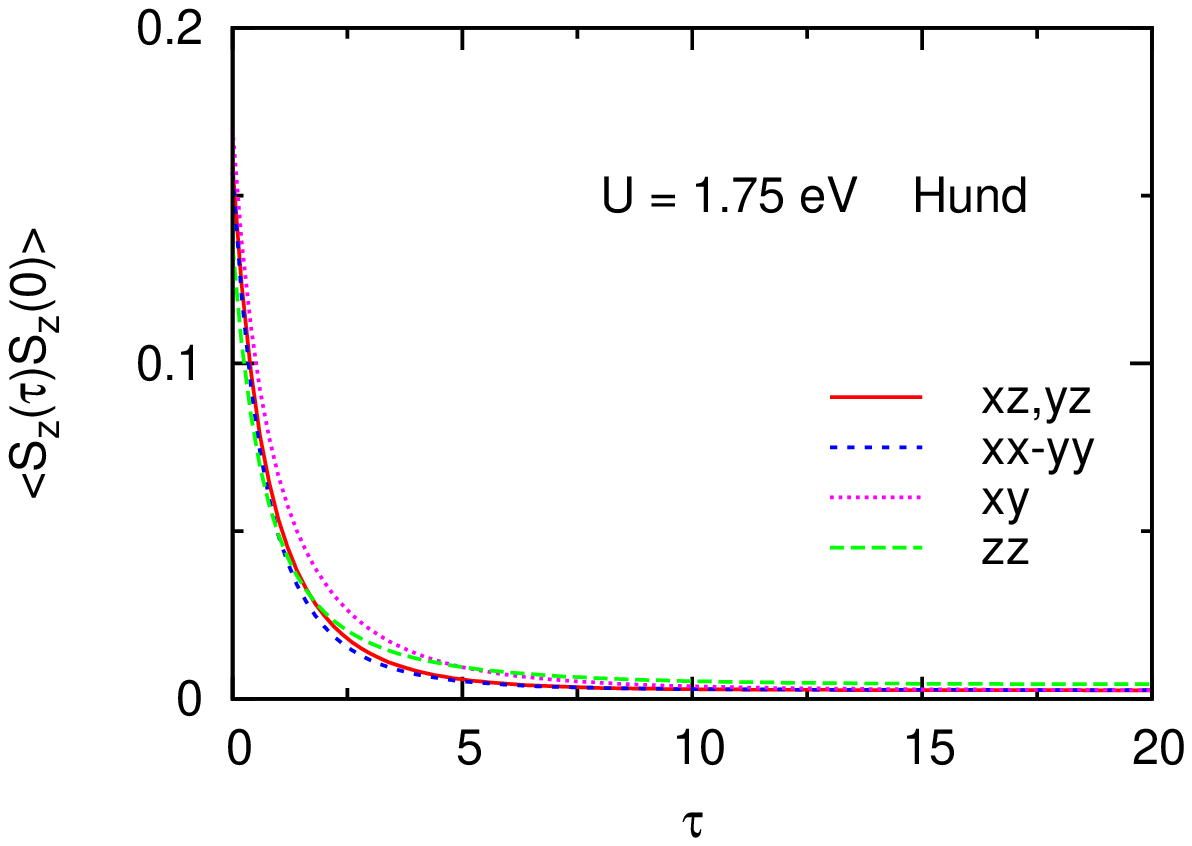}
\end{center}
\vskip-7mm \ \ \ (a)\hfill  \mbox{\hskip5mm}
\begin{center}
\includegraphics[width=4.5cm,height=6.5cm,angle=-90]{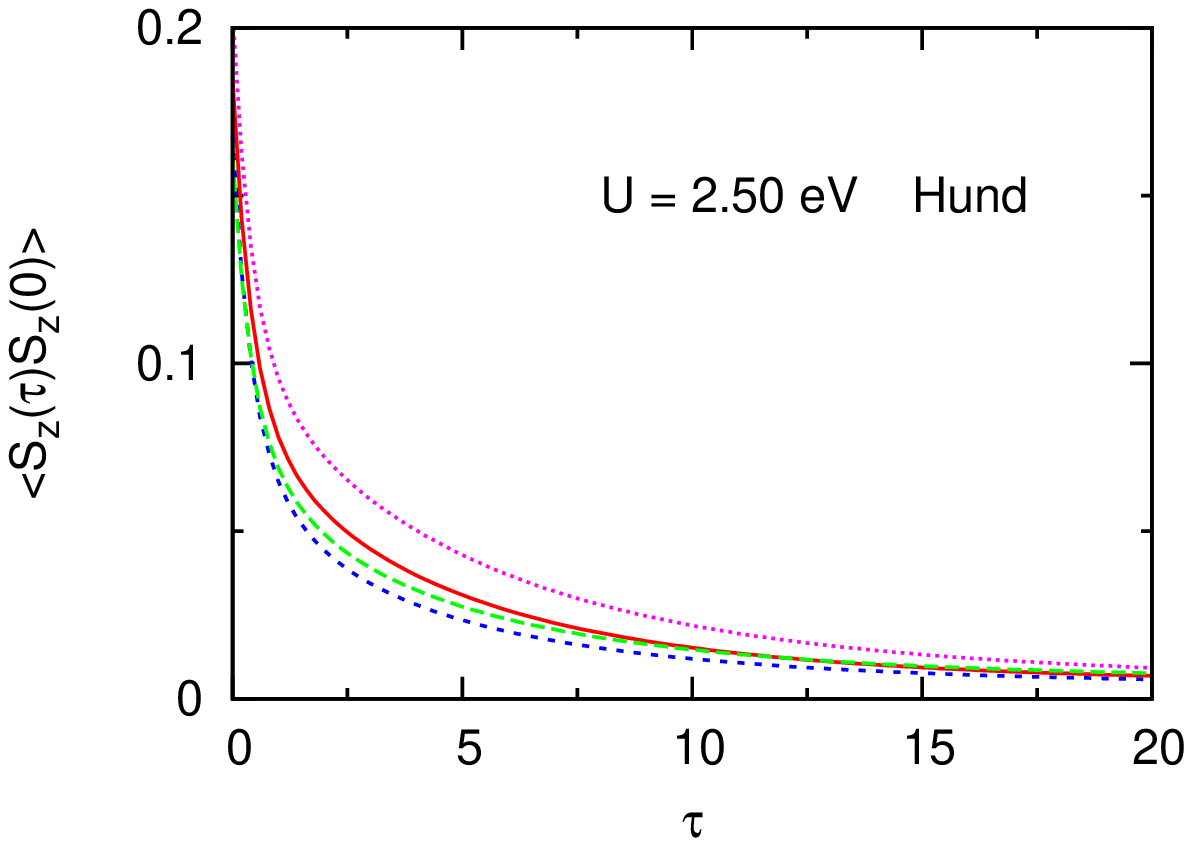}
\end{center}
\vskip-7mm \ \ \ (b)\hfill  \mbox{\hskip5mm}
\begin{center}
\includegraphics[width=4.5cm,height=6.5cm,angle=-90]{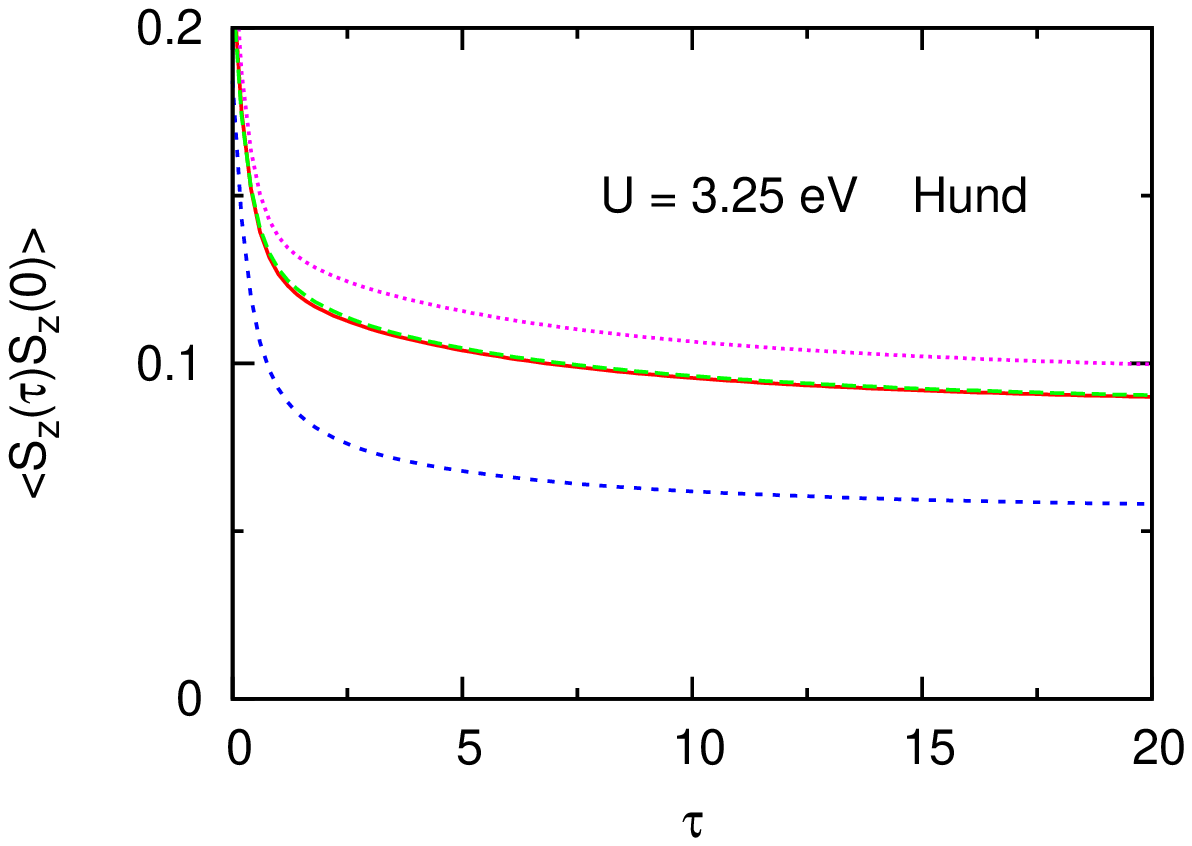}
\end{center}
\vskip-7mm \ \ \ (c)\hfill \mbox{\hskip5mm}
\vskip-1mm
\caption{(Color online)
Variation of orbital dependent spin-spin correlation functions 
with imaginary time $\tau$, for three Coulomb energies and Hund
coupling. (a) $U=1.75$~eV,  (b) $U=2.5$~eV, (c) $U=3.25$~eV. 
}\label{SzSzH}\end{figure}

\begin{figure}  [t!] %7
\begin{center}
\includegraphics[width=4.5cm,height=6.5cm,angle=-90]{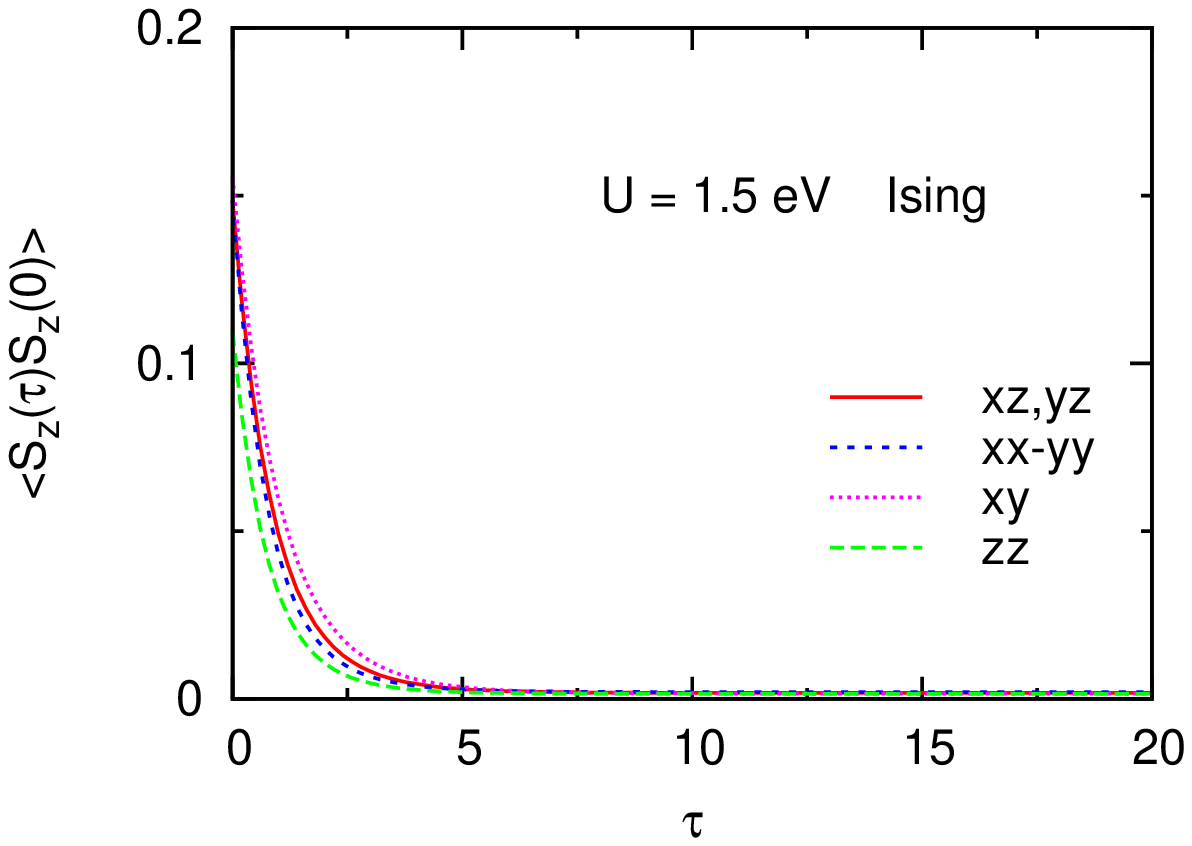}
\end{center}
\vskip-7mm \ \ \ (a)\hfill \mbox{\hskip5mm}
\begin{center}
\includegraphics[width=4.5cm,height=6.5cm,angle=-90]{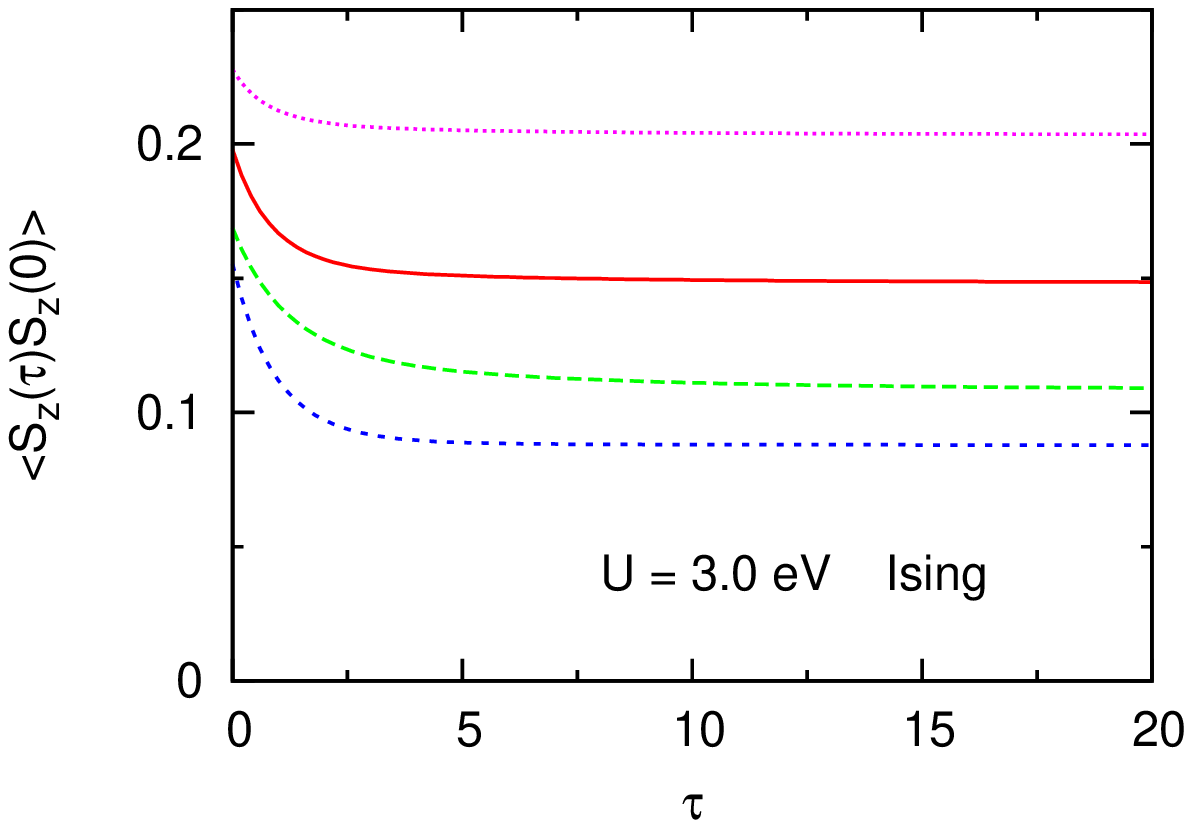}
\end{center}
\vskip-7mm \ \ \ (b)\hfill \mbox{\hskip5mm}
\vskip-1mm
\caption{(Color online)
Variation of orbital dependent spin-spin correlation functions 
with imaginary time $\tau$, for two Coulomb energies and Ising
exchange. (a) $U=1.5$~eV,  (b) $U=3.0$~eV; $J=U/4$.
}\label{SzSzI}\end{figure}

To explore the origin of the Fermi-liquid to non-Fermi-liquid
transition we have evaluated the spin-spin correlation
function $C_{mz}(\tau)=\langle S_{mz}(\tau)S_{mz}(0)\rangle$, where 
$\tau$ denotes the imaginary time. Fig.~\ref{SzSzH} shows these 
orbital dependent functions for several Coulomb energies and Hund
coupling. At low values of $U$, the $C_{mz}(\tau)$ decay to zero, 
as expected for a Fermi liquid. The orbital components of the spin 
susceptibility \begin{equation}
     \chi_m\sim \int_0^\beta d\tau\, \langle S_{mz}(\tau)S_{mz}(0)\rangle 
\end{equation}
are then independent of temperature, indicating Pauli behavior.
With increasing $U$, the decay becomes less rapid and finite values 
are approached at large $\tau$ (for $\tau\ll\beta=1/T$), demonstrating  
the formation of local moments $S_m$ simultaneously in all subbands.
The susceptibility components are then proportional to $\beta$, so that 
$\chi_m\sim S_m(S_m+1)/T$, as expected for Curie-Weiss behavior.    
As shown in Fig.~\ref{SzSzI}, for Ising coupling the formation of 
frozen moments sets in at much lower Coulomb energies.      

A similar spin-freezing transition was recently found by Werner {\it et al.}
\cite{werner} for a fully degenerate three-band model near $n=2$ occupancy. 
Using continuous-time QMC as impurity solver, the paramagnetic 
$U-n$ phase diagram was shown to exhibit Fermi-liquid properties 
at small $U$. For increasing $U$ and $n>1.5$, an incoherent metallic 
phase with local moments appears, which is then replaced by 
a Mott insulating phases at integer occupancies $n=2$ and $n=3$.  
Beyond the critical value of $U$, the low-frequency limit of the 
self-energy exhibits a finite onset of similar magnitude as shown 
here in Fig.~\ref{SigvsU}. In the present five-band study, we find 
in addition that this transition changes approximately from continuous
to first-order when Hund exchange is replaced by Ising-like coupling. 

\begin{figure}  [t!] %8
\begin{center}
\includegraphics[width=4.5cm,height=6.5cm,angle=-90]{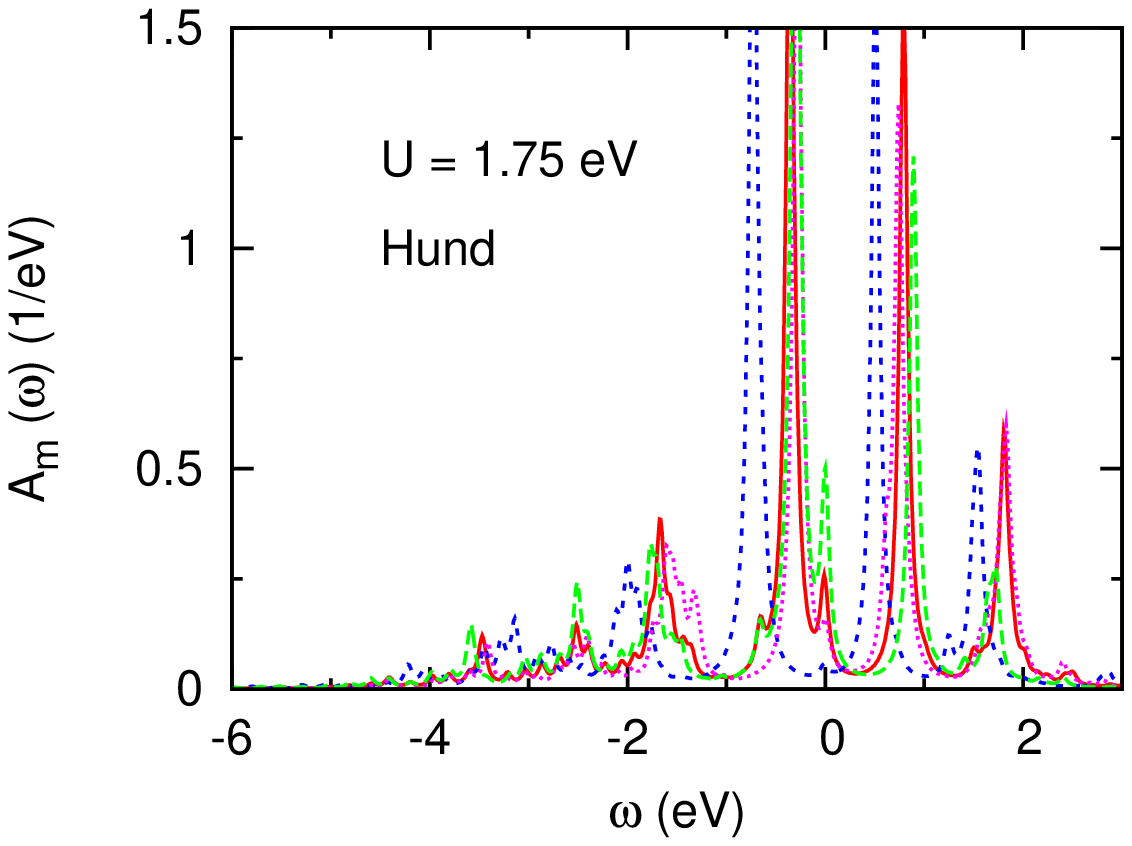}
\end{center}
\vskip-7mm \ \ \ (a)\hfill \mbox{\hskip5mm}
\begin{center}
\includegraphics[width=4.5cm,height=6.5cm,angle=-90]{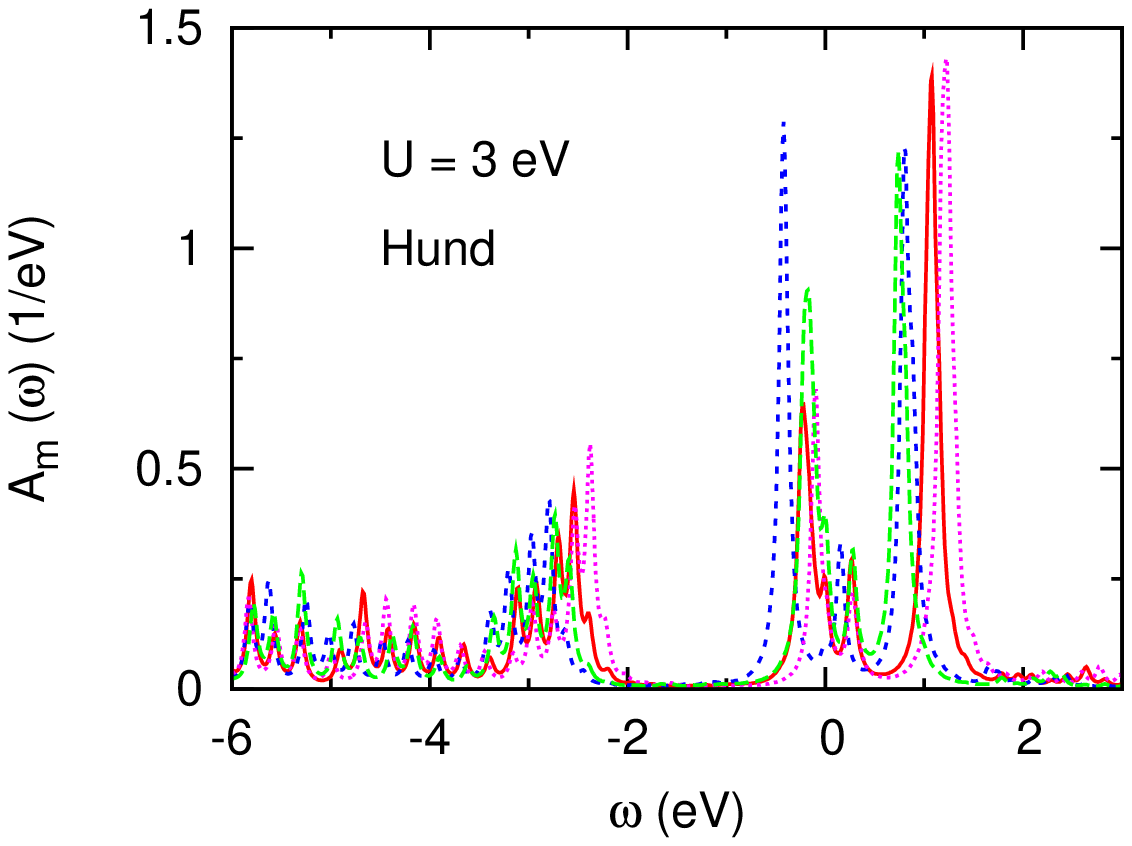}
\end{center}
\vskip-7mm \ \ \ (b)\hfill \mbox{\hskip5mm}
\begin{center}
\includegraphics[width=4.5cm,height=6.5cm,angle=-90]{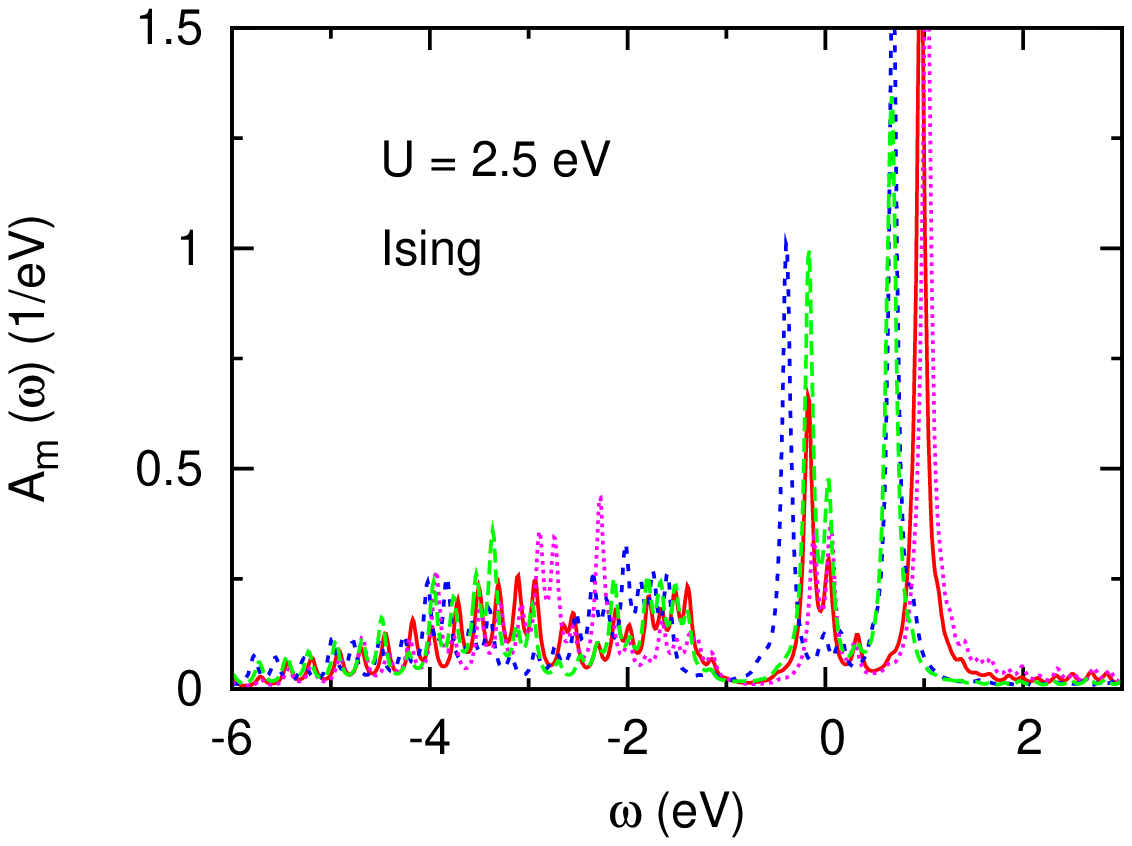}
\end{center}
\vskip-7mm \ \ \ (c)\hfill \mbox{\hskip5mm}
\vskip-1mm
\caption{(Color online)
Spectral distributions of Fe $3d$ subbands for (a) $U=1.75$~eV,
(b) $U=3$~eV; Hund coupling; (c) $U=2.5$~eV: Ising exchange; 
with $J=U/4$ and $0.05$~eV broadening.
}\label{Am}\end{figure}

To illustrate the correlation-induced transfer of spectral weight 
in LaFeAsO we show in Figs.~\ref{Am}(a) and (b) the $3d$ 
spectral distributions for two Coulomb energies and Hund coupling. 
For simplicity we plot here the ED cluster spectra since they do not 
require analytic continuation to real frequencies. The main effect 
at $U=1.75$~eV is the band narrowing both below and above $E_F$. 
In addition, spectral weight is shifted below the bottom of the 
$3d$ bands, indicating the formation of weak lower Hubbard bands. 
At $U=3$~eV, spectral weight in the occupied part of the $3d$ bands 
is greatly reduced and the Hubbard bands are much more prominent. 
Since there is little experimental evidence for any significant 
lower Hubbard bands, the above results imply that $U$ within the 
present 5-band description should be approximately $2.0\ldots2.5$~eV,
where the precise value depends on the magnitude of $J$. 
The unoccupied states are also shifted closer to $E_F$, but there 
is no evidence of any upper Hubbard peaks. Note that the lower 
Hubbard bands are very broad. This is related to large uncorrelated 
band width and to the multiplet structure induced by Hund's rule coupling.      
The spectra for Ising exchange shown in panel (c) are qualitatively
similar. Many small differences arise because of the absence of 
spin flip and pair hopping exchange processes. 
The different multiplet structures associated with Hund and Ising
coupling will be considered in more detail elsewhere.      

The above results demonstrate the importance of a proper treatment of 
exchange interactions. For instance, if because of $pd$ screening 
realistic values of $U$ and $J$ for LaFeAsO are approximately 
$2.0\ldots2.5$~eV and $J\approx U/4$, respectively, full Hund coupling 
suggests that the system is moderately strongly correlated, with $3d$ 
effective mass enhancements of the order of $m^*/m = 1/Z_m\approx 2\ldots3$.
In contrast, if spin-flip and pair-exchange processes are ignored 
(as is usually done in Hirsch-Fye QMC calculations to avoid sign 
problems, e.g., in Refs.~\cite{anisimov,shorikov,skornyakov}), 
the same Coulomb and exchange parameters suggest that system has crossed 
the boundary towards non-Fermi-liquid behavior, with strongly reduced 
lifetimes of electronic states close to $E_F$. It would be interesting 
to inquire whether a $dp$ formulation of $H({\bf k})$ yields a similar 
qualitative difference between Hund and Ising exchange.

At sufficiently large Coulomb energies, $n=6$ integer occupancy should
eventually lead to a Mott insulating phase. We have increased $U$ up to
$6$~eV while keeping $J=0.75$~eV constant. Both for Hund and Ising 
coupling, the system evolves towards an orbital selective phase, 
where the $d_{xz,yz,xy}$ subset is either in or close to an insulating
phase and the $d_{x^2-y^2,z^2}$ subset remains in the strongly incoherent 
metallic state.\cite{comment} 
Thus, for realistic Coulomb and exchange we conclude that in the present 
five-orbital ED/DMFT description the system is far below the $n=6$ Mott 
insulating region. Orbital selective phases in five-band systems at 
$n=6$ occupancy, with subbands split via a crystal field, were also 
found in Refs.\cite{luca,lauchli}        

We close this subsection by pointing out that several papers have recently 
discussed the role of coexisting itinerant and localized electrons in iron 
pnictides.\cite{si1,wu,luca,kou,hackl,si2}  
A simple system which exhibits this kind of coexistence is the half-filled 
two-band model where the narrow band is Mott localized while the wide band 
is still metallic.\cite{prb04,prl05,costi,biermann}
Coulomb interactions between these two types of electrons give rise to 
bad-metallic behavior in the itinerant band, with a finite scattering rate
and a pseudogap at $E_F$ in the case of Ising exchange,\cite{prb04} and 
marginal Fermi-liquid behavior in the case of full Hund coupling. 
\cite{biermann,costi}         
As we have seen above, the important feature of FeAsLaO at $n=6$ occupancy 
is that the spin freezing transition occurs simultaneously in all five $d$
bands, far below the Mott transition. Thus, in the range of Coulomb and 
exchange energies of interest, all subbands have similar occupancies, 
effective masses, and low-frequency scattering rates.
The subbands therefore do not split into itinerant and localized subsets.
Only at much larger $U$, the $d_{xz,yz,xy}$  and $d_{x^2-y^2,z^2}$ bands 
gradually tend towards $0.5$ and $0.75$ occupancy.
   
\subsection{Doped LaFeAsO: correspondence between pnictides and cuprates}

\begin{figure}  [t!] %9
\begin{center}
\includegraphics[width=4.5cm,height=6.5cm,angle=-90]{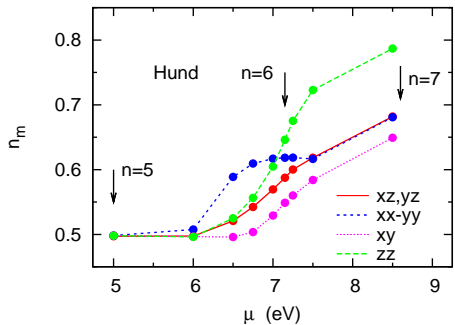}
\end{center}\vskip-3mm
\caption{(Color online)
Fe $3d$ orbital occupancies as functions of chemical potential for 
$U=2.5$~eV with $J=0.625$~eV and Hund coupling. The arrows indicate 
the values of $\mu$ associated with integer total occupancies. 
}\label{nvsmu}\end{figure}

The results discussed above are for paramagnetic, undoped LaFeAsO, with Fe 
$3d$ occupancy $n=6$. To illustrate the effect of electron and hole doping 
we show in Fig.~\ref{nvsmu} the orbital occupancies as functions of chemical 
potential at fixed $U=2.5$~eV. In order to elucidate the doping variation 
of many-body properties, the one-electron Hamiltonian is kept unchanged. 
The doping range extends from one hole to one electron. (At half-filling 
$n=5$, the chemical potential is $\mu=4.5 U - 10 J=5$~eV.) Evidently, the 
degree of orbital polarization depends strongly on the total occupancy. 
For $n=5$, all $d$ bands become half-filled and some orbital components 
Im\,$\Sigma_m(i\omega_n)$ are porportional to $1/\omega_n$. The spectral 
distributions reveal that the system then is a Mott insulator where all 
$d$ bands are split into lower and upper Hubbard peaks.  

In the case of electron doping, all orbital occupancies increase in a 
similar fashion. At $n=7$, the self-energies (not shown) indicate that 
the system is a weakly correlated Fermi liquid. Even for Ising exchange, 
the quasi-particle weights are in the range $Z_m=0.26\ldots0.56$. 
These properties differ strikingly from those discussed above at $n=6$, 
where under the same interaction conditions the system is much closer 
to bad-metallic behavior. Thus, the Fermi-liquid to non-Fermi-liquid   
phase boundary for $n=7$ is shifted to larger values of $U$. 
These results suggest a fundamental asymmetry of LaFeAsO with respect
to doping. For hole doping, bad-metallic behavior should increase, 
while electron doping reinforces Fermi-liquid properties.   
This behavior is consistent with results for the three-band model
\cite{werner} where the Fermi-liquid to non-Fermi-liquid phase boundary 
was also found to shift to larger $U$ as the occupancy moves farther 
away from half-filling. 

\begin{figure}  [b!] %10
\begin{center}
\includegraphics[width=4.5cm,height=6.5cm,angle=-90]{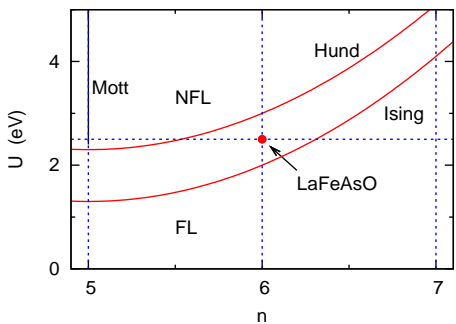}
\end{center}\vskip-3mm
\caption{(Color online)
Schematic phase diagram for doped LaFeAsO. Red curves: 
boundaries between Fermi-liquid (FL) and non-Fermi-liquid (NFL) phases 
for Hund and Ising exchange. At half-filling ($n=5$) a Mott insulating 
phase exists down to rather small $U$. The undoped material at $n=6$ 
with moderate $U\approx2.5$~eV is a Fermi liquid for Hund coupling, 
but an incoherent metal for Ising exchange.   
The $n=6$ Mott phase is located at $U>6$~eV.
}\label{phase}\end{figure}

On the basis of the above results we obtain the paramagnetic $U-n$ 
phase diagram shown in Fig.~\ref{phase}, where, for $U=2.5$~eV and 
$J=U/4$ Hund exchange, undoped LaFeAsO lies just below the Fermi-liquid 
/ non-Fermi-liquid phase boundary. Smaller $U$ and larger $J$ would 
move this point farther below this phase boundary. In the limit of 
one-hole doping, the system is a Mott insulator, whereas, for electron 
doping, Fermi-liquid properties dominate. For Ising exchange, the 
phase boundary is shifted to roughly 1~eV lower $U$ values, so that 
$U$ would have to be less than $\sim 2$~eV to preserve Fermi-liquid 
behavior.  

Of course, in real LaFeAsO, this phase diagram should  be more complicated 
because of the inevitable modification of the 
one-electron properties with doping\cite{mazin2,xu}
and because of the antiferromagnetic phase\cite{cruz} observed at $n=6$.
Nevertheless, the paramagnetic limit permits to draw an interesting 
analogy between the present multi-band iron pnictide and the single-band 
two-dimensional Hubbard model that is frequently used to investigate 
Coulomb correlations in high-$T_c$ cuprates. 

\begin{figure}  [t!] %11
\begin{center}
\includegraphics[width=4.5cm,height=6.5cm,angle=-90]{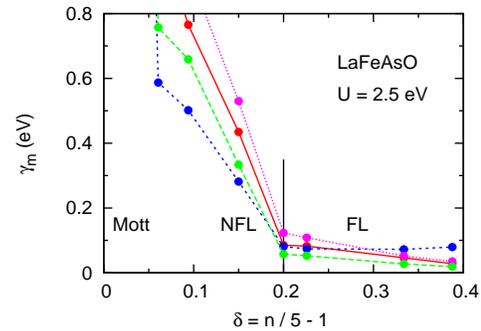}
\end{center}
\vskip-7mm \ \ \ (a)\hfill \mbox{\hskip5mm}
\begin{center}
\includegraphics[width=4.5cm,height=6.5cm,angle=-90]{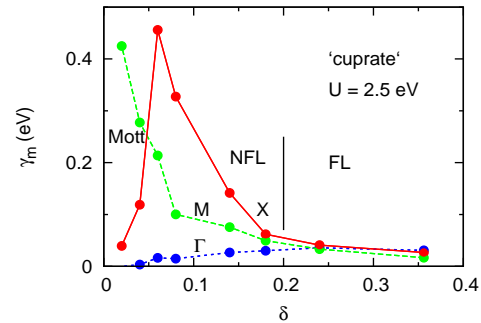}
\end{center}
\vskip-7mm \ \ \ (b)\hfill \mbox{\hskip5mm}
\vskip-1mm
\caption{(Color online)
(a) Orbital-dependent scattering rates $\gamma_m$ for LaFeAsO 
as functions of doping relative to half-filling, $\delta=n/5-1$ 
(Hund exchange, $J=U/4$).  
(b) Cluster components of scattering rate for single-band
two-dimensional Hubbard model as functions of hole doping (nearest and 
next-nearest neighbor hopping $t=0.25$~eV and $t'=-0.075$~eV, 
respectively, $U=2.5$~eV, $T=0.01$~eV).\cite{prb09} 
The vertical bars denote the approximate doping concentrations of
the Fermi-liquid to non-Fermi-liquid transition.  
}\label{gvsd}\end{figure}

Fig.~\ref{gvsd}(a) shows the orbital components of the low-frequency 
scattering rate $\gamma_m\approx -{\rm Im}\,\Sigma_m(i\omega_0)$ for 
LaFeAsO as a function of electron doping relative to half-filling, i.e.,
$\delta=n/5-1$, where $n=6$ for the undoped material. As discussed above, 
in the limit of one-hole doping ($n=5$), the system is a Mott insulator, 
while for $\delta>0.2$ it becomes a weakly correlated Fermi liquid. 
In the intermediate region, for $\delta<0.2$, the scattering rate 
increases sharply so that the system is dominated by non-Fermi-liquid 
properties due to the formation of frozen moments. Thus, close to $n=6$ 
the system is near the Fermi-liquid to non-Fermi-liquid phase boundary. 

As shown in panel (b), this behavior is remarkably similar to the one 
found within cluster ED/DMFT for hole doping in the two-dimensional 
Hubbard model.\cite{prb09} Analogous results have been obtained by 
several groups.\cite{jarrell2001,senechal,parcollet,park,vidhya,werner2} 
This model yields a Mott insulator at half-filling and exhibits a 
non-Fermi-liquid pseudogap phase up to a critical doping 
$\delta_c \approx 0.15\ldots 0.2$. Ordinary metallic behavior is 
restored in the overdoped region, $\delta>\delta_c$. 
For electron doping the results are similar, except for a smaller 
critical doping which marks the onset of bad-metallic behavior. 
In view of the analogy seen in Fig.~\ref{gvsd}, the Mott phase that 
is relevant for LaFeAsO is not the one that should eventually appear 
at $n=6$ for large $U$, but the one that exists at realistic values of 
$U$ and $J$ at $n=5$ occupancy.

The results discussed above demonstrate that the Mott transition in
multi-orbital and multi-site Hubbard models is far more complex than
the paramagnetic metal to insulator transition obtained in single-band,
single-site DMFT calculations. The presence of interactions channels
involving orbitals or sites not only affects the overall magnitude of the
critical Coulomb energy, but gives rise to a much richer phase diagram. 
In particular, as we have seen here for LaFeASO, a frozen-moment,  
non-Fermi-liquid phase emerges between the Fermi-liquid and insulating 
regions. Conceptually, this non-Fermi-liquid behavior is closely related 
to the pseudogap phase in hole-doped cuprates which arises from short-range     
Coulomb correlations.

\section{Summary}

Multi-band ED/DMFT has been used to investigate the effect of correlations 
in the iron pnictide LaFeAsO. Starting from an accurate five-band 
tight-binding single-particle Hamiltonian, the many-body properties 
are evaluated by extending single-site ED/DMFT to five orbitals, each
hybridizing with two bath levels.
This scheme is particularly useful at very low temperatures, large Coulomb 
energies and for fully rotationally invariant Hund's rule coupling. 
It is shown that correlation effects in LaFeAsO give rise to a 
paramagnetic transition from a Fermi-liquid phase to a non-Fermi-liquid 
phase at a critical Coulomb energy ($pd$ screened) of about $3$~eV.
This transition appears to be continuous and is caused by the formation of 
Fe $3d$ local moments. Below this transition, the quasi-particle weight of 
all $d$ orbitals is strongly reduced and orbital polarization is less 
pronounced than in the uncorrelated density functional band structure. 
A Mott insulating phase does not appear below $U=6$~eV.       
For Ising exchange the Fermi-liquid to non-Fermi-liquid transition
also exists, but the critical value of $U$ is shifted to about $2$~eV. 
Moreover, the transition appears to be first order rather than continuous. 

Despite the multi-band nature of LaFeAsO and the important role of exchange 
interactions, this system exhibits an interesting relationship to the 
single-band two-dimensional Hubbard model if the doping concentration is 
defined with respect to the half-filled $3d$ shell. Within this picture, 
iron pnictides and cuprates exhibit the same sequence of paramagnetic phases 
with increasing doping: from Mott insulator to bad metal to Fermi liquid. 
Thus, fluctuations between orbitals or lattice sites lead to similar physics.
According to the available experimental evidence, LaFeAsO is located on 
the weakly to moderately correlated side of the Fermi-liquid to 
non-Fermi-liquid phase boundary. 
Thus, iron pnictide materials appear to have two parent compounds: the 
anti-ferromagnetic semi-metal at $n=6$ and the Mott insulator at $n=5$.

\bigskip
{\bf Acknowledgements:}  
A. L. likes to thank Igor Mazin, Luca de' Medici and Massimo Capone 
for comments. The work of H. I. is supported by the Grant-in-Aid for 
Scientific Research (Grant No. 20540191) and by the Nihon University
Strategic Projects for Academic Research.
A. L.'s calculations were carried out on the J\"ulich Jump computer.


\begin{thebibliography}{99}
\bibitem{comment}
   In Ref.~\cite{haule1}, a Mott insulator was found at $U=4.5$~eV, $J=0.7$~eV,
   while in Ref.~\cite{shorikov} even $U=5$~eV did not yield a Mott phase.
\end{thebibliography}
\end{document}